\documentclass[a4paper,twocolumn,aps,superscriptaddress,prb,10pt]{revtex4-1}
\usepackage{graphicx}
\usepackage{color}
\usepackage{tikz}
\usepackage{pgfplots} 
\usepackage{amssymb}
\pgfplotsset{tick scale binop=\times}

\begin{document}
\title{Fractional charge by fixed-node diffusion Monte Carlo}
\author{Matej Ditte}
\affiliation{Department of Physics, University of Ostrava, 30. dubna 22, 701 03 Ostrava, Czech Republic}
\author{Mat\'u\v{s} Dubeck\'{y}}\email{matus.dubecky@osu.cz}
\affiliation{Department of Physics, University of Ostrava, 30. dubna 22, 701 03 Ostrava, Czech Republic}
\affiliation{ATRI, Slovak University of Technology, Paul\'inska 16, 917 24 Trnava, Slovakia}

\date{\today}

\begin{abstract}
Fixed-node diffusion Monte Carlo (FNDMC)  is a stochastic quantum many-body method that has a great potential in electronic structure theory.
We examine how FNDMC satisfies exact constraints, linearity and derivative discontinuity of total energy $E(N)$ vs. fractional electron number $N$, if combined with mean-field trial wave functions that miss such features. H and Cl atoms with fractional charge reveal that FNDMC is well able to restore the piecewise linearity of $E(N)$. The method uses ensemble and projector ingredients to achieve the correct charge localization. Water-solvated Cl$^-$ complex illustrates superior performance of FNDMC for charged noncovalent systems.
\end{abstract}
\maketitle

{\it Motivation:} fixed-node diffusion Monte Carlo (FNDMC) is a many-body stochastic quantum Monte Carlo (QMC) projector method\cite{Anderson1975,Reynolds1982,Umrigar1993,Mitas1991} that has been popular in electronic structure theory for its accuracy, scalability and  versatility\cite{Kolorenc2011rev,Luchow2011rev,Austin2012rev,Morales2014rev,Dubecky2016rev}.
FNDMC acts in a continuum position space; for a Hamiltonian $\hat H$ it projects out an exact ground state (GS) $\Psi$ that has non-zero overlap with the antisymmetric trial state  $\Psi_T$, in imaginary \mbox{time $\tau$:}
\begin{equation}
\Psi= \lim_{\tau\rightarrow\infty} \exp[-\tau(\hat H - E_T)]\Psi_T,
\end{equation}
where $E_T$ is an offset energy that keeps the norm of $\Psi$ asymptotically constant. The total FNDMC energy is an upper bound to the exact energy\cite{Moskowitz1982} and the related FN bias goes quadratically with the nodal displacement error\cite{Mitas1991}.
$\Psi_T$ can be well sophisticated so that the related FN bias becomes negligible\cite{Morales2012,Clay2015}, nevertheless, practical FNDMC feasible for large systems requires simple and efficient ans\"atze like the popular Slater-Jastrow wave function \cite{Ceperley1977},
    $\Psi_T=\Psi_S J$,
where $\Psi_\mathrm{S}$ is a single Slater determinant and $J$ is a positive-definite explicit correlation Jastrow term\cite{Jastrow1955}. 

Mean-field theories used to produce $\Psi_S$, including DFT\cite{Per2012}, as well as the states $\Psi_S$ and $\Psi_T$ (see below), however miss fundamental constraints on exact electronic structure theory: the total energy $E(N)$ as a continuous function of a particle number $N$ must show piecewise linear relationship with possible derivative discontinuities at integer $N$\cite{Perdew1982,Yang2000,Mori2009}. Lack thereof has been referred to as (de)localization error and it has severe consequences; it leads to an artificial charge adjustment and related spurious energy minimization which causes poor predictivity of charge-transfer/reaction barriers, band gaps, or noncovalent interactions (e.g., solvation of ions)\cite{Cohen2008,Mori2014rev}. In addition, $E(N)$ must also satisfy constancy condition of fractional spin \cite{Mori2009} important for correct description of strong correlation. 

Here we examine how FNDMC satisfies linearity of $E(N)$ for fractional $N$, and,  derivative discontinuity at integer $N$, if combined with $\Psi_T$ based on popular spin-restricted mean-field $\Psi_S$ that show unphysical convex $E(N)$ dependency\cite{Mori2014rev}. Such an analysis is important, for instance, to understand if the method is able to provide right energetics of charged noncovalent systems\cite{Cohen2008} for right reasons. We consider atoms with fractional charge, H and Cl, to show that FNDMC is well able to restore the piecewise linearity of $E(N)$ from the states that do not possess such a property. Walker population analyzes indicate that ensemble and projector features are both operative in this achievement. Insights gained from water-solvated Cl$^-$ complex suggest that accuracy and robustness of FNDMC in charge-involving noncovalent interactions relate to the accurate charge localization.

{\it Theory:} the effective fractional charge per atom was achieved for both, H and Cl, by modeling cube (system) composed of 8 atoms (subsystems) of the same type separated enough (infinite-separation limit) so that the interactions between the subsystems can be neglected. The Hamiltonian of such a system with $M=\sum_{i=1}^8N_i$ electrons,
\begin{equation}
 \hat H = \sum_{i=1}^{8}\hat H^i,
\end{equation}
satisfies the Schr\"odinger equation,
\begin{equation}
 \hat H \Psi_k(M) = E(M) \Psi_k(M),
\end{equation}
where
\begin{equation}
 \Psi_k(M)=\hat A[\prod_{i=1}^8\Psi^i_{\alpha_i}(N_i)]
\end{equation}
is a \mbox{$k$-th} $g(M)$-times degenerate antisymmetric GS wave function $[g(M)=\prod_{i=1}^3 g_i(N_i)]$ assembled as an antisymmetrized product of individual subsystem $g_i(N_i)$-times degenerate wave functions $\Psi^i_{\alpha_i}(N_i)$ that each satisfies
\begin{equation}
 \hat H^i \Psi^i_\alpha(N_i) = E^i (N_i) \Psi^i{_{\alpha_i}}(N_i),
\end{equation}
$\hat A$ is an $M$-electron antisymmetrization operator, and,
\begin{equation}
   E(M)=\sum_{i=1}^8 E^i(N_i)
\end{equation}
is a corresponding degenerate total energy of the system. Addition of an integer charge $q$ to such a system adds effective charge $q/8$ per subsystem thus enabling fractional charge as well. 

QMC simulations of systems in large separation limit require additional considerations for the lack of VMC thermalization within usual simulation times (see Methods), and, non-ergodicity (electron localization on specific subsystems) due to the branching term in FNDMC that eliminates configurations with small weights, including those with electrons possibly attempting to diffuse between the distant subsystems. We argue that the total energy expectation values $\langle E\rangle$ from such FNDMC simulations are energetically correct.

One  needs to realize that within a properly initialized non-ergodic equilibrium FNDMC simulation (starting from $\Psi_T^2$), any walker \mbox{$\mathbf{R}_j=(\mathbf{r}_1,\mathbf{r}_2,\mathbf{r}_3,\dots,\mathbf{r}_N)_j$}, irrespective of its specific localization $\{N_i\}_j$, i.e. a specific set of occupation numbers of the subsystems labeled by $i$, randomly samples one of the degenerate symmetry-broken GS wave functions $\Psi_k$ and contributes to $\langle E \rangle$ according to the so-called mixed distribution $\Psi_k\Psi_T$. For a totally symmetric $\Psi_T$, like, e.g., our $\Psi_T$ based on a spin-restricted Slater determinant, that satisfies $\langle\Psi_T|\Psi_k\rangle\neq 0$ for all $k$, it is not hard to accept that the population-based $\langle E\rangle$, that is of our sole interest here, is not affected by the symmetry breaking.

{\it Hydrogen cube:} first, we present the results obtained for H$_8$ cube\cite{Mori2014rev} with an edge of $a=$1000~\AA{}, 0-16 electrons and singlet spin multiplicity ($N_\alpha=N_\beta$), allowing us to study essentially isolated H atom with an effective fractional charge in range $0\leq N\leq 2$, i.e., along \mbox{H$^+\rightarrow$H$\rightarrow $H$^{-}$} pathway and 0.25$e$ increments (Fig.~1). In agreement with the previous fractional-charge studies of H\cite{Mori2009,Mori2014,Mori2014rev}, we find that the considered mean-field methods, HF and DFT with PBE\cite{Perdew1996} functional, miss linearity and derivative discontinuity of $E(N)$ and show unphysical convex behavior instead (Fig.~1, red, blue).
\begin{figure}[b!]
 \centering
\includegraphics[bb=0 0 1620 1266,width=200pt]{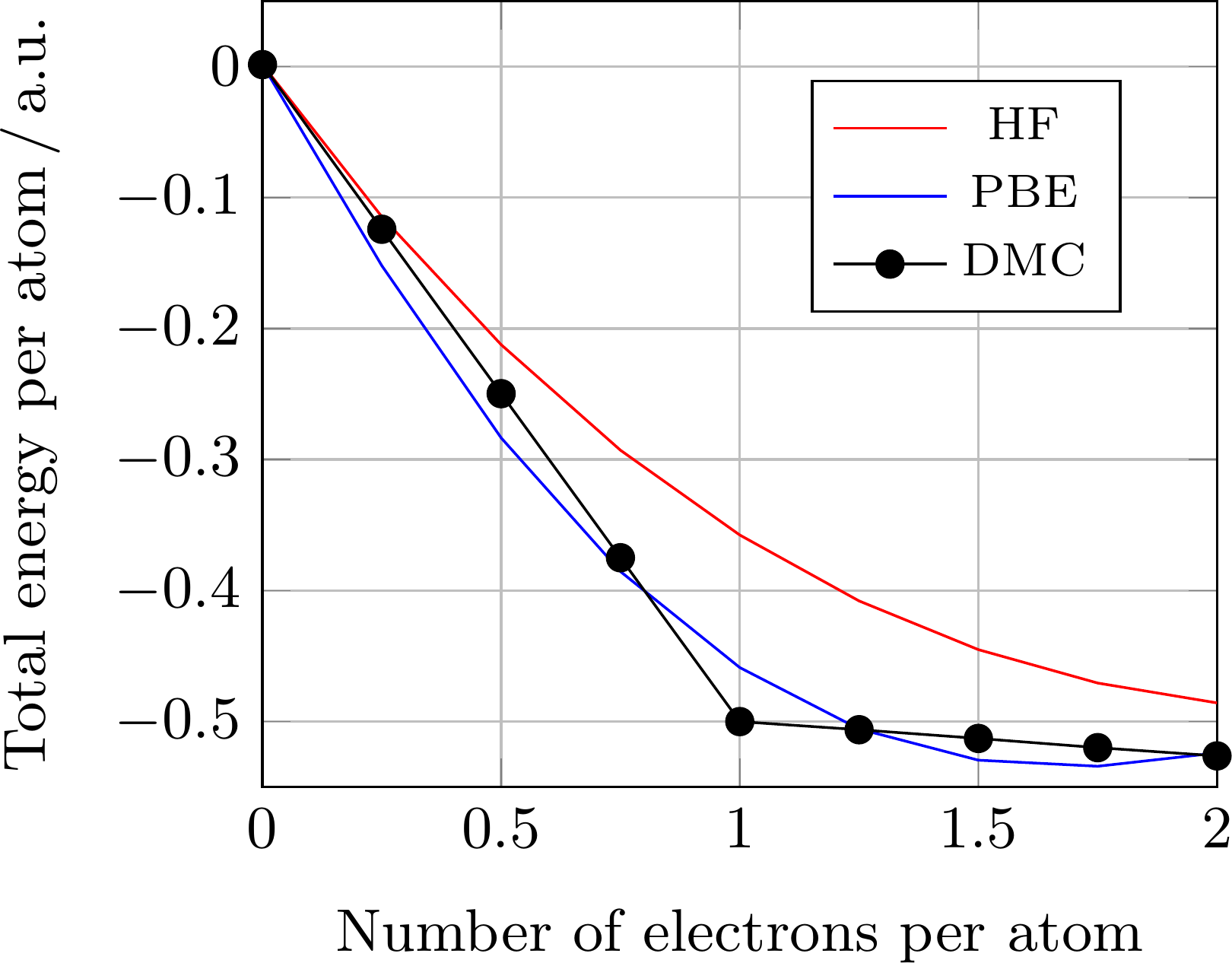}
 \label{fig_H}
 \caption{Hydrogen cube: total energy $E(N)$ per atom vs. electron occupation number $N$ per atom obtained by mean-field (HF, PBE) and FNDMC (DMC) methods. FNDMC error bars (not shown) are smaller than the symbol size.}
\end{figure}
Interestingly, FNDMC with related $\Psi_T$ (cf. Fig.~3) recovered piecewise linear $E(N)$ dependency within the statistical resolution (Fig.~1, black). The results were indistinguishable for HF and PBE (more $\Psi_T$ types were thus not considered). FNDMC with $\Psi_T$ based on PBE orbitals produced electron affinity EA(H)=0.713(1)~eV to be compared vs. experimental value of 0.754~eV\cite{Lykke1991}.
Slight discrepancy of $\sim$0.04~eV is attributed to the use of effective core potential (ECP) and residual FN bias due to 1-determinant\cite{Gasperich2017}.
Note that if FNDMC method satisfies linearity condition, the quality of the nodal surface determines only the slope of the actual $E(N)$ segment.

{\it Chlorine cube:} next we consider Cl$_8$ cube ($a=$1000~\AA{}) in order to examine how FNDMC describes fractional charge in many-electron system. 
We consider only singlet states ($N_\alpha=N_\beta$).
Effective fractional charge per Cl atom ranges between $16\leq N\leq 18$ (\mbox{$^1$Cl$^+\rightarrow^2$Cl$\rightarrow ^1$Cl$^{-}$}) with 0.25$e$ increments (Fig.~2). We observe that although the mean-field methods (HF, PBE, HSE06\cite{Heyd2006}) miss linearity, and derivative discontinuity\cite{Cohen2008} at point $N=17$ per atom (neutral Cl), FNDMC well recovers both features from the related $\Psi_T$ and produces accurate \mbox{EA(Cl)=3.64(2)~eV} consistent with experiment (\mbox{3.613~eV\cite{Berzinsh1995}}). It appears that FNDMC is well able to produce correct energetics in fractional-charge systems. 
\begin{figure}[!h]
 \centering
  \includegraphics[bb=0 0 1638 1266,width=200pt]{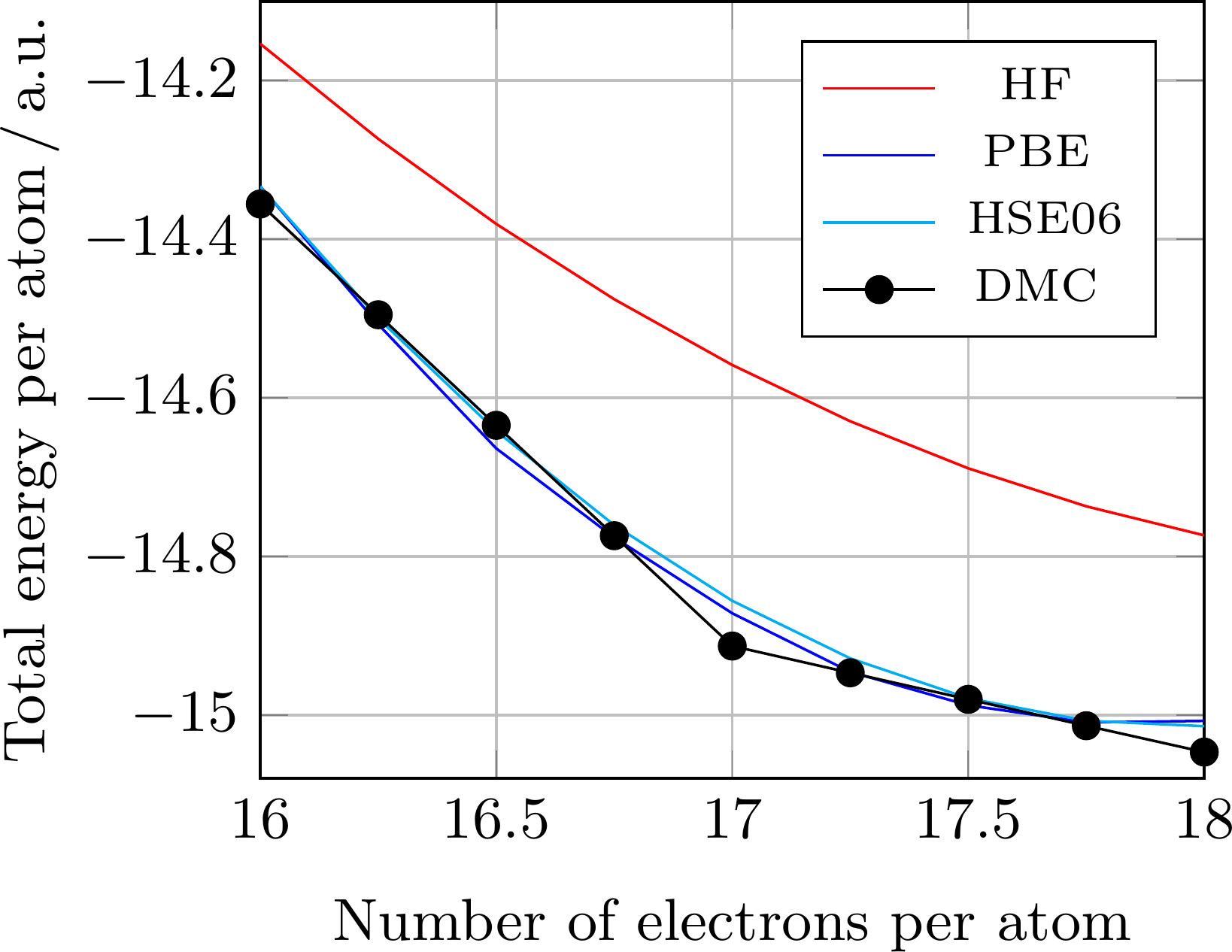}
 \label{fig_Cl}
  \caption{Chlorine cube: total energy $E(N)$ per atom vs. electron occupation number $N$ per atom obtained by mean-field (HF, PBE, HSE06) and FNDMC (DMC) methods. FNDMC error bars (not shown) are smaller than the symbol size.}
\end{figure}

{\it Wave function analysis:} let us now focus on how FNDMC achieves linearity of $E(N)$ in fractional charge models.
As mentioned above, the electrons within each walker localize on specific centers, leading to the correspondence \mbox{$\mathbf{R}_j\leftrightarrow\{N_i\}_j$} between the explicit (alive) walker position (varying along simulation) and subsystem occupation vector in occupation representation (fixed during the non-ergodic simulation unless the walker dies out). Since each walker randomly samples one of the degenerate GS wave functions $\Psi_k$, and the subsystem occupations randomly vary between the walkers, we find it convenient to introduce irreducible subsystem occupation vector $\vec\lambda_j$ obtained as a descending-ordered set of subsystem occupations $\{N_i\}_j$, \mbox{$\vec\lambda_j=(\mathrm{max}\{N_i\}_j,\dots,\mathrm{min}\{N_i\}_j)$}. In addition, we introduce a corresponding vector of total subsystem spin momenta $\vec \sigma_j$ ordered in a descending way for each subset of identical $\lambda_k$.
The pair $(\lambda_k,\sigma_k)_j$, labels the state of a subsystem $k$ (after reordering) with $\lambda_k$ electrons and a total spin momentum $\sigma_k$, as sampled by the walker $j$. We arrive at the more convenient correspondence, \mbox{$\mathbf{R}_j\leftrightarrow|\vec\lambda,\vec\sigma\rangle_j$},
that enables equal-footing comparisons between the walkers irrespective of the specific localization (and local position), and, unambiguous analysis of the system state within FNDMC simulation. We are interested in coefficients $\{c_{\vec\lambda,\vec\sigma}\}$ of an expansion of irreducible GS
\begin{equation}
 \Psi(\mathbf{R},\tau)=\sum_{\vec\lambda,\vec\sigma}c_{\vec\lambda,\vec\sigma}|\vec\lambda,\vec\sigma\rangle,
\end{equation}
subject to $\sum_{\vec\lambda,\vec\sigma} |c_{\vec\lambda,\vec\sigma}|^2=1$, that could be obtained from a stochastic realization of $\Psi(\mathbf{R},\tau)$ sampled by $K$ walkers within equilibrium QMC simulation,
\begin{equation}
 \Psi(\mathbf{R},\tau)=\sum_{j=1}^K\delta[\mathbf{R}-\mathbf{R}_j(\tau)],
\end{equation}
by virtue of the projectors,
\begin{equation}
 \Psi(\mathbf{R},\tau)=\sum_{\vec\lambda,\vec\sigma}\sum_{j=1}^K |\vec\lambda,\vec\sigma\rangle\langle\vec\lambda,\vec\sigma|\delta_{\mathbf{R},\mathbf{R}_j(\tau)}\rangle.
\end{equation}

\begin{figure}[!b]
 \centering
\includegraphics[width=200pt,bb=0 0 1620 1266]{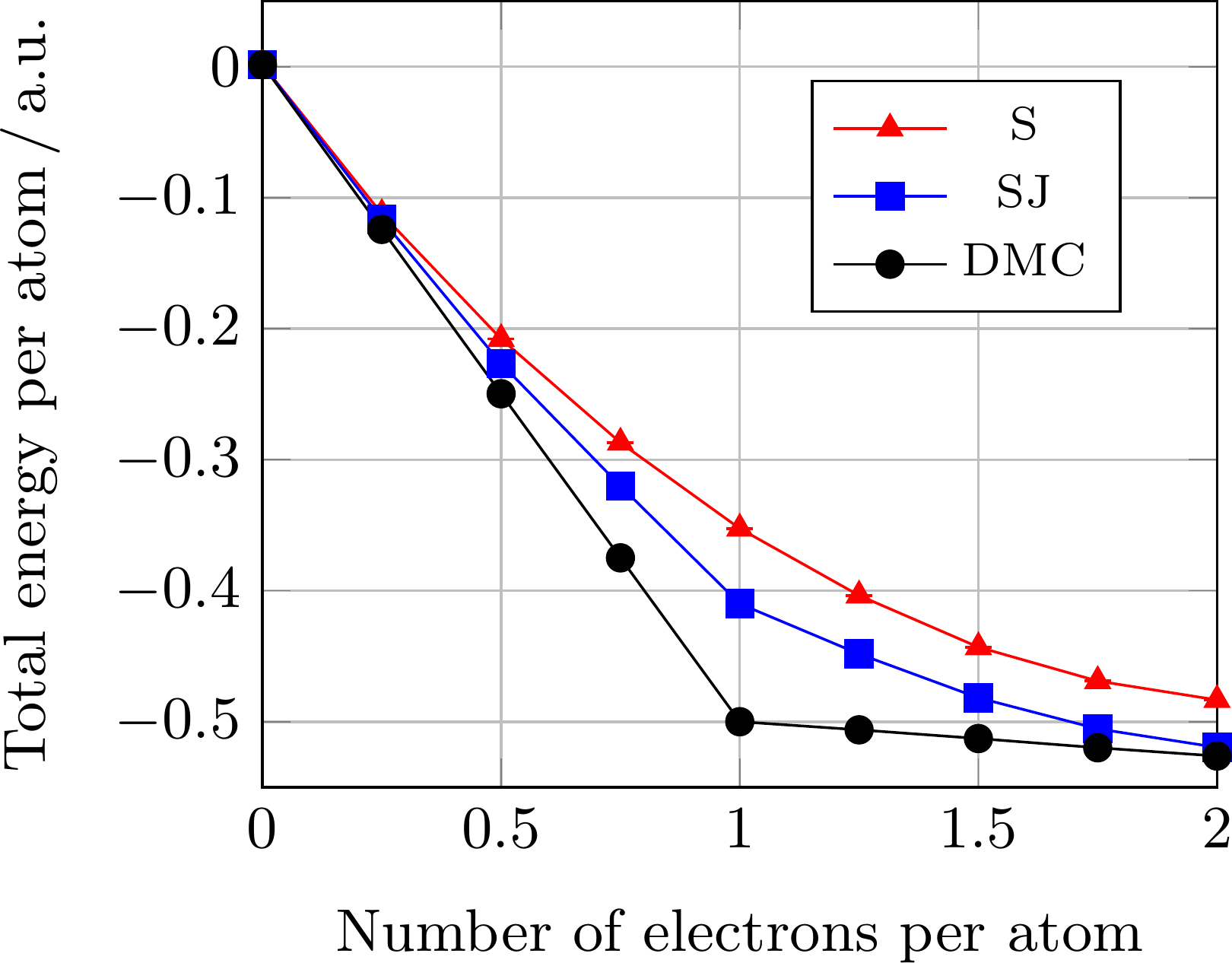}
 \label{fig_H}
 \caption{Hydrogen cube: total energy $E(N)$ per atom vs. electron number per atom from VMC with $\Psi_S$ (S) and Slater-Jastrow $\Psi_T$ (SJ) using PBE orbitals, and, FNDMC (DMC).}
\end{figure}

For $M$ electrons in the system with a specific partition $\vec\lambda_j$ between the subsystems, the GS consistent with the subsystem total energy convexity condition\cite{Perdew1982},
\begin{equation}
E^i(N)\leq \frac{1}{2} [E^i(N+1)+E^i(N-1)],
\end{equation}
requires that each distant subsystem contains \mbox{$a=\mathrm{int}(M/8)$} electrons, and, for nonzero $b=M-8a$, $b$ subsystems contain 1 additional electron each. $\vec\sigma$ is such that it minimizes the total energy.
For $M=10$, any 2 of the 8 subsystems must contain 2 electrons and the remaining subsystems 1 electron each, $\vec\lambda=(2,2,1,1,1,1,1,1)$, assuming a spin GS, e.g., $\vec\sigma=(0,0,0.5,0.5,0.5,-0.5,-0.5,-0.5)$ for a singlet H$_8$ cube. Other partitions $\vec\lambda$ (or non-GS $\vec\sigma$) represent excited-state configurations, e.g., $\vec\lambda=(2,2,2,1,1,1,1,0)$, or, $\vec\lambda=(3,1,1,1,1,1,1,1)$. Therefore, for the special case of distant subsystems, an exact GS may be conveniently identified by 1-term eq.~7 expansions.

In Fig.~3, we compare the total energies $E(N)$ obtained for H cube by VMC sampling of $|\Psi_{\mathrm{S}}^2|$ and $|\Psi_T^2|$, and FNDMC using $\Psi_T$. Clearly, contrary to the piecewise linear FNDMC, the VMC total energies show convex behavior, although in case of SJ, it seems that the derivative discontinuity starts to develop due to the use of Jastrow term\cite{Goetz2017}. For $\Psi_{\mathrm{S}}$, the $E(N)$ curve is indeed consistent with HF in Fig.~1 within the statistical error.

Analysis of the related walker ensembles in terms of $|\vec\lambda,\vec\sigma\rangle$ configurations revealed  multi-configuration states containing large fractions of excited-states for VMC, 
and, only GS configurations for FNDMC equilibrium ensembles, for all $N$, orbital types, and Cl cube as well (cf. Fig.~5, Appendix I). For instance, $\Psi_S$/$\Psi_T$ of PBE orbitals of neutral H cube with $N=1$ electrons contained only 6\%/11\% of GS configurations and 6/5 types of configurations, whereas $\Psi$ sampled by FNDMC using $\Psi_T$ contained GS configurations only. For $N=1.25$, $\Psi_S$/$\Psi_T$ contained 22\%/38\% of GS configurations and 6/4 configuration types, whereas, again, FNDMC contained 100\% of GS. Details of such analyzes for H and Cl cubes, DFT and HF orbitals, VMC and FNDMC, and all considered $N$, are reported in Appendix III.

The observation that FNDMC produces 100\% of GS configurations in our fractional-charge models is very important as it may improve our understanding of the method robustness in charged\cite{Wineman-Fisher2019} and charge-transfer systems (e.g.,  reactions\cite{Swann2017,Krongchon2017}, excited states\cite{Valsson2010,Dubecky2011}). It has been shown\cite{Yang2000} that the linearity of $E(N)$ for fractional $N$, and thus accurate charge localization, directly relates to the subsystem GS occupations. Namely, if the total energy of the system can be expressed, using \mbox{$E^i(\lambda_i)=E^j(\lambda_i)$}, as
\begin{equation}
    E(M)=(8-b)E^1(a)+bE^1(a+1),
\end{equation}
where $a$ and $b$ were defined above, then $E(N)=E(M)/8$ for $N=M/8$ is a piecewise linear function.
The success of FNDMC in recovering the $E(N)$ piecewise linearity is thus attributed to the ensemble nature of the method causing that each electron resides on a specific subsystem at a time (fractional density is only possible via ensemble), and, projector property that apparently projects out proper (GS) configurations.

{\it Cl$^-$(H$_2$O)$_6$:} Finally, we consider a practical application of the concepts discussed above for noncovalent interactions of charged systems\cite{Wineman-Fisher2019}. It was found that FNDMC provides correct $E(N)$ dependency for Cl cube contrary to mean-field (Fig.~2). In systems with a charged Cl atom, one would thus generally expect that FNDMC would be more robust than, say, DFT, and insensitive to the $\Psi_T$ orbitals. We have considered interaction energies\cite{Dubecky2016rev} ($\Delta E$) of Cl$^-$(H$_2$O)$_6$ complex by various DFT approximations and FNDMC using the same DFT functionals. Fig.~4 summarizes the results and illustrates superior robustness of FNDMC that reproduces the reference CCSD(T)/complete-basis-set results to within 1\% for all $\Psi_T$. In contrast, $\Delta E$ from DFT and HF altogether varies by more than 20\% due to the known (de)localization error. 

This is now easy to understand. Randomly chosen DFT determines fractional charge $N$ from a certain interval according to its intrinsic delocalization error. In addition, it produces energy on a convex curve with its intrinsic curvature, that  varies between functionals\cite{Cohen2006}. As a consequence, for a large set of DFT approximations, one obtains a large distribution of biases in the total energy of a complex, and, consequently in $\Delta E$, as observed. On the other hand, if FNDMC predicts a correct charge localization, as expected from the results reported above, the fractional charge $N$ in subsystems of a complex is accurate and well defined. Inaccuracy of $\Psi_T$ dictates only the slope of the linear dependency, which leaves only a limited interval of bias that can be realized with varying (reasonable) $\Psi_T$. The dispersion of the results is thus much smaller, or, in other words, the method is much more robust, as observed, and we attribute its success to the accurate charge localization.
\begin{figure}[!t]
\begin{tabular}{cc}
\includegraphics[width=250pt,bb=0 0 2102 1326]{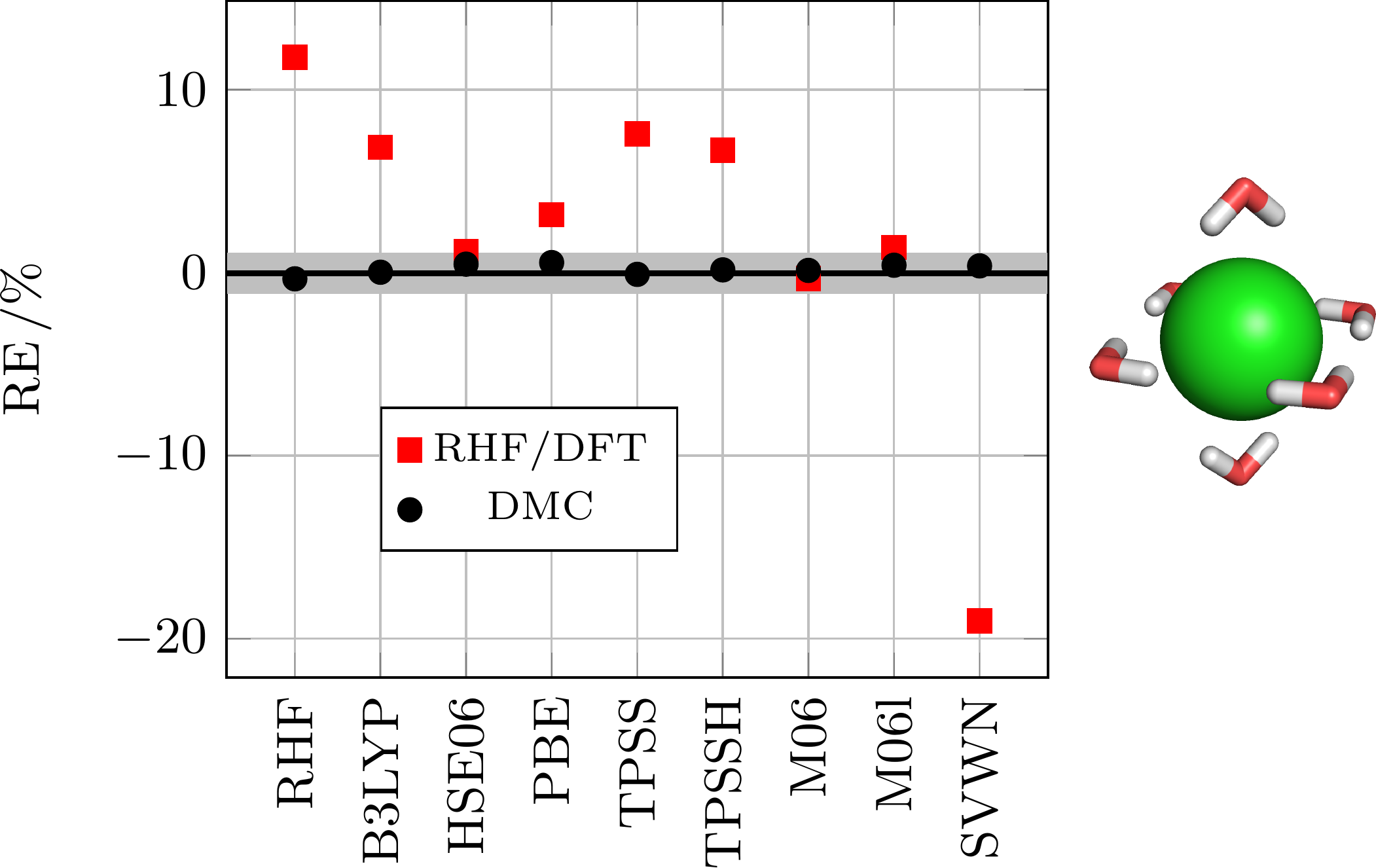}
\end{tabular}
\label{fig_ClWa}
 \caption{Relatative errors (RE) of the interaction energies for solvated anion Cl$^-$(H$_2$O)$_6$ complex (depicted) computed by RHF, DFT and DMC using $\Psi_T$ based on the same orbitals, compared vs. the CCSD(T)/CBS reference. Gray region indicates benchmark level (1\%). FNDMC error bars (not shown) are smaller than the symbol size.}
\end{figure}

{\it Summary:} Fractional-charge computations of H and Cl atoms revealed that FNDMC is well able to restore the correct piecewise linear $E(N)$ behavior from $\Psi_T$ that does not have such a property. Insights from the walker population analyses, in combination with formal discussions, indicate that FNDMC correctly describes $E(N)$ linearity thanks to its ensemble nature and projection property. Inaccuracy of $\Psi_T$ is expected to determine only the slope of $E(N)$ segments. This finding enables understanding of FNDMC robustness, as demonstrated in noncovalent Cl$^-$(H$_2$O)$_6$ complex, from a new perspective. Study of fractional spin by FNDMC is underway.

{\it Methods:} {\small
%
Naive simulations of cube models with an edge of $a=1000$~\AA{} would not be possible because the VMC equilibration times would vastly overcome the maximum simulation times available. In order to overcome this issue in an automated fashion (avoiding manual initialization of configurations that may be biased by user imagination), the walkers for large-cube FNDMC simulations were initialized by ergodic VMC thermalization and stepwise quasi-adiabatic extension of the system size starting from small $a$. We verified that the walker distributions produced in this way sample $\Psi_T^2$, and, in case of a Slater determinant with Hartree-Fock (HF) orbitals, VMC reproduced the HF total energy for both, H and Cl, and all considered $N$. This was not the case for walkers initialized directly in a large cube.

FNDMC computations used QMCPACK\cite{Kim2018} code with imaginary time step of 0.005 a.u. and T-moves\cite{Casula2006}.  Nuclei were represented by Burkatzki-Filippi-Dolg effective core potentials\cite{Burkatzki2008}. Single-determinant Slater-Jastrow $\Psi_T$ were based on orbitals expanded in 1s-augmented VTZ one-particle basis sets without highest angular momentum channels\cite{Dubecky2017b} obtained with a tight SCF convergence (Gaussian G09\cite{G09}). Jastrows containing up to electron-electron-nucleus terms, with a cutoff radius of 10 Bohr, were optimized by the linear method\cite{Toulouse2007}, as usual, but cube simulations reused Jastrows optimized for H$^-$ and Cl, respectively. The structure of Cl$^-$(H$_2$O)$_6$ (reported in Appendix~II) was optimized at the MP2/aug-VTZ level (G09).
}

{\small
~~~Discussions with A. Gendiar have been fruitful. Financial support by Czech Science Foundation (\mbox{18-24321Y}, \mbox{18-25128S}), University of Ostrava (IRP201826), and ERDF (ITMS 26220220179), is gratefully acknowledged. The computations used IT4Innovations National Supercomputing Center (LM2015070).
}

~\newline
\newpage
\section*{Appendix}
\section{Chlorine cube}~
\begin{figure}[h!]
 \centering
\includegraphics[width=200pt,bb=0 0 1638 1266]{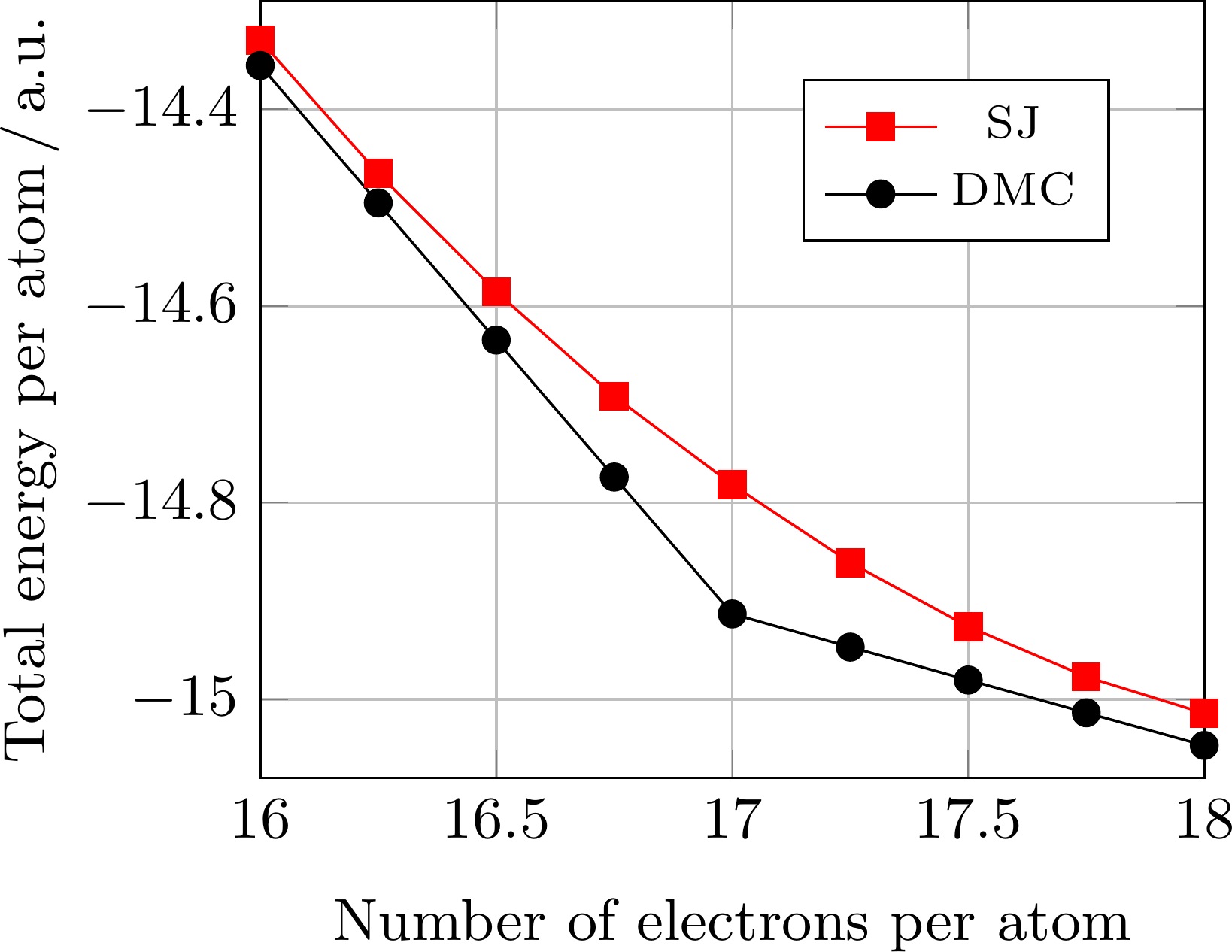}
 \label{fig_H}
 \caption{Chlorine cube: total energy $E(N)$ per atom vs. electron occupation number $N$ per atom obtained by by VMC sampling of local energy from Slater-Jastrow (SJ) $\Psi_T$ using Slater determinant of HSE06 orbitals, and, FNDMC (DMC) using the same SJ $\Psi_T$.}
\end{figure}

{
\section{Structure of solvated chlorine-anion complex}
The structure of the considered Cl$^-$(H$_2$O)$_6$ complex optimized at the MP2/aug-VTZ level in G09 (xyz-format):
\begin{verbatim}
       19
 
Cl     0.000000     0.000000     0.000000
 O     3.290138     0.000000     0.000000
 O    -3.290138     0.000000     0.000000
 O     0.000000     3.264842     0.000000
 O     0.000000    -3.264842     0.000000
 O     0.000000     0.000000     3.353781
 O     0.000000     0.000000    -3.353781
 H     2.664747     0.734646     0.000000
 H     2.664747    -0.734646     0.000000
 H    -2.664747     0.734646     0.000000
 H    -2.664747    -0.734646     0.000000
 H     0.000000     2.636302     0.731619
 H     0.000000     2.636302    -0.731619
 H     0.000000    -2.636302     0.731619
 H     0.000000    -2.636302    -0.731619
 H     0.000000     0.732695     2.727551
 H     0.000000    -0.732695     2.727551
 H     0.000000     0.732695    -2.727551
 H     0.000000    -0.732695    -2.727551
\end{verbatim}
}
\newpage
~

\newpage
{
\onecolumngrid
\section{Analysis of walkers}
Walker ensemble analyzes for H and Cl cube models, various methods (VMC, FNDMC), and, trial states (Slater determinant $\Psi_S$, Slater-Jastrow $\Psi_T$). For each $N$ (in electrons per atom), reported are \%, $\vec\lambda$ and $\vec\sigma$. GS configurations are indicated by the bold font.

\begin{table}[h!] 
\centering 
H-cube, VMC with $\Psi_S$ of PBE orbitals \\ 
\begin{tabular}{lll}
0.25 electrons per atom  & & \\ 
\textbf{87.86\%} & \textbf{(1,1,0,0,0,0,0,0)} & \textbf{(0.5,-0.5,0,0,0,0,0,0)} \\ 
12.14\% & (2,0,0,0,0,0,0,0) & (0,0,0,0,0,0,0,0) \\\\ 
0.5 electrons per atom  & & \\ 
\textbf{56.27\%} & \textbf{(1,1,1,1,0,0,0,0)} & \textbf{(0.5,0.5,-0.5,-0.5,0,0,0,0)} \\ 
37.81\% & (2,1,1,0,0,0,0,0) & (0,0.5,-0.5,0,0,0,0,0) \\ 
5.92\% & (2,2,0,0,0,0,0,0) & (0,0,0,0,0,0,0,0) \\\\ 
0.75 electrons per atom  & & \\ 
46.39\% & (2,1,1,1,1,0,0,0) & (0,0.5,0.5,-0.5,-0.5,0,0,0) \\ 
28.59\% & (2,2,1,1,0,0,0,0) & (0,0,0.5,-0.5,0,0,0,0) \\ 
\textbf{21.92\%} & \textbf{(1,1,1,1,1,1,0,0)} & \textbf{(0.5,0.5,0.5,-0.5,-0.5,-0.5,0,0)} \\ 
3.1\% & (2,2,2,0,0,0,0,0) & (0,0,0,0,0,0,0,0) \\\\ 
1.0 electrons per atom  & & \\ 
37.57\% & (2,2,1,1,1,1,0,0) & (0,0,0.5,0.5,-0.5,-0.5,0,0) \\ 
25.28\% & (2,1,1,1,1,1,1,0) & (0,0.5,0.5,0.5,-0.5,-0.5,-0.5,0) \\ 
24.54\% & (2,2,2,1,1,0,0,0) & (0,0,0,0.5,-0.5,0,0,0) \\ 
6.44\% & (2,2,2,2,0,0,0,0) & (0,0,0,0,0,0,0,0) \\ 
\textbf{6.16\%} & \textbf{(1,1,1,1,1,1,1,1)} & \textbf{(0.5,0.5,0.5,0.5,-0.5,-0.5,-0.5,-0.5)} \\ 
0.01\% & (3,2,1,1,1,0,0,0) & (0.5,0,0.5,-0.5,-0.5,0,0,0) \\\\ 
1.25 electrons per atom  & & \\ 
46.58\% & (2,2,2,1,1,1,1,0) & (0,0,0,0.5,0.5,-0.5,-0.5,0) \\ 
28.34\% & (2,2,2,2,1,1,0,0) & (0,0,0,0,0.5,-0.5,0,0) \\ 
\textbf{21.76\%} & \textbf{(2,2,1,1,1,1,1,1)} & \textbf{(0,0,0.5,0.5,0.5,-0.5,-0.5,-0.5)} \\ 
3.3\% & (2,2,2,2,2,0,0,0) & (0,0,0,0,0,0,0,0) \\ 
0.01\% & (2,2,2,1,1,1,1,0) & (1.0,0,0,0.5,-0.5,-0.5,-0.5,0) \\ 
\text{0.01\%} & \text{(2,2,1,1,1,1,1,1)} & \text{(0,-1.0,0.5,0.5,0.5,0.5,-0.5,-0.5)} \\\\ 
1.5 electrons per atom  & & \\ 
\textbf{56.62\%} & \textbf{(2,2,2,2,1,1,1,1)} & \textbf{(0,0,0,0,0.5,0.5,-0.5,-0.5)} \\ 
37.17\% & (2,2,2,2,2,1,1,0) & (0,0,0,0,0,0.5,-0.5,0) \\ 
6.2\% & (2,2,2,2,2,2,0,0) & (0,0,0,0,0,0,0,0) \\\\ 
1.75 electrons per atom  & & \\ 
\textbf{87.61\%} & \textbf{(2,2,2,2,2,2,1,1)} & \textbf{(0,0,0,0,0,0,0.5,-0.5)} \\ 
12.35\% & (2,2,2,2,2,2,2,0) & (0,0,0,0,0,0,0,0) \\ 
0.02\% & (3,2,2,2,2,1,1,1) & (0.5,0,0,0,0,0.5,-0.5,-0.5) \\ 
0.02\% & (3,2,2,2,2,1,1,1) & (-0.5,0,0,0,0,0.5,0.5,-0.5) \\\\ 
2.0 electrons per atom  & & \\ 
\textbf{99.58\%} & \textbf{(2,2,2,2,2,2,2,2)} & \textbf{(0,0,0,0,0,0,0,0)} \\ 
0.24\% & (3,2,2,2,2,2,2,1) & (0.5,0,0,0,0,0,0,-0.5) \\ 
0.17\% & (3,2,2,2,2,2,2,1) & (-0.5,0,0,0,0,0,0,0.5) \\\\ 
\end{tabular} 
\label{tableHcubePBEeqnoj} 
\end{table}

\begin{table}[h!] 
\centering 
H-cube, VMC with $\Psi_S$ of HF orbitals \\ 
\begin{tabular}{lll}
0.25 electrons per atom  & & \\ 
\textbf{87.57\%} & \textbf{(1,1,0,0,0,0,0,0)} & \textbf{(0.5,-0.5,0,0,0,0,0,0)} \\ 
12.43\% & (2,0,0,0,0,0,0,0) & (0,0,0,0,0,0,0,0) \\\\ 
0.5 electrons per atom  & & \\ 
\textbf{56.26\%} & \textbf{(1,1,1,1,0,0,0,0)} & \textbf{(0.5,0.5,-0.5,-0.5,0,0,0,0)} \\ 
37.27\% & (2,1,1,0,0,0,0,0) & (0,0.5,-0.5,0,0,0,0,0) \\ 
6.47\% & (2,2,0,0,0,0,0,0) & (0,0,0,0,0,0,0,0) \\\\ 
0.75 electrons per atom  & & \\ 
46.58\% & (2,1,1,1,1,0,0,0) & (0,0.5,0.5,-0.5,-0.5,0,0,0) \\ 
28.25\% & (2,2,1,1,0,0,0,0) & (0,0,0.5,-0.5,0,0,0,0) \\ 
\textbf{21.91\%} & \textbf{(1,1,1,1,1,1,0,0)} & \textbf{(0.5,0.5,0.5,-0.5,-0.5,-0.5,0,0)} \\ 
3.26\% & (2,2,2,0,0,0,0,0) & (0,0,0,0,0,0,0,0) \\\\ 
1.0 electrons per atom  & & \\ 
38.43\% & (2,2,1,1,1,1,0,0) & (0,0,0.5,0.5,-0.5,-0.5,0,0) \\ 
24.73\% & (2,1,1,1,1,1,1,0) & (0,0.5,0.5,0.5,-0.5,-0.5,-0.5,0) \\ 
24.46\% & (2,2,2,1,1,0,0,0) & (0,0,0,0.5,-0.5,0,0,0) \\ 
6.26\% & (2,2,2,2,0,0,0,0) & (0,0,0,0,0,0,0,0) \\ 
\textbf{6.12\%} & \textbf{(1,1,1,1,1,1,1,1)} & \textbf{(0.5,0.5,0.5,0.5,-0.5,-0.5,-0.5,-0.5)} \\\\ 
1.25 electrons per atom  & & \\ 
47.28\% & (2,2,2,1,1,1,1,0) & (0,0,0,0.5,0.5,-0.5,-0.5,0) \\ 
27.47\% & (2,2,2,2,1,1,0,0) & (0,0,0,0,0.5,-0.5,0,0) \\ 
\textbf{22.39\%} & \textbf{(2,2,1,1,1,1,1,1)} & \textbf{(0,0,0.5,0.5,0.5,-0.5,-0.5,-0.5)} \\ 
2.87\% & (2,2,2,2,2,0,0,0) & (0,0,0,0,0,0,0,0) \\\\ 
1.5 electrons per atom  & & \\ 
\textbf{56.19\%} & \textbf{(2,2,2,2,1,1,1,1)} & \textbf{(0,0,0,0,0.5,0.5,-0.5,-0.5)} \\ 
37.51\% & (2,2,2,2,2,1,1,0) & (0,0,0,0,0,0.5,-0.5,0) \\ 
6.29\% & (2,2,2,2,2,2,0,0) & (0,0,0,0,0,0,0,0) \\ 
0.01\% & (3,2,2,2,1,1,1,0) & (-0.5,0,0,0,0.5,0.5,-0.5,0) \\\\ 
1.75 electrons per atom  & & \\ 
\textbf{87.51\%} & \textbf{(2,2,2,2,2,2,1,1)} & \textbf{(0,0,0,0,0,0,0.5,-0.5)} \\ 
12.49\% & (2,2,2,2,2,2,2,0) & (0,0,0,0,0,0,0,0) \\\\ 
2.0 electrons per atom  & & \\ 
\textbf{99.93\%} & \textbf{(2,2,2,2,2,2,2,2)} & \textbf{(0,0,0,0,0,0,0,0)} \\ 
0.04\% & (3,2,2,2,2,2,2,1) & (0.5,0,0,0,0,0,0,-0.5) \\ 
0.04\% & (3,2,2,2,2,2,2,1) & (-0.5,0,0,0,0,0,0,0.5) \\\\ 
\end{tabular} 
\label{tableHcubeRHFeqnoj} 
\end{table}

\begin{table}[h!] 
\centering 
H-cube, VMC with $\Psi_T=\Psi_S J$ using $\Psi_S$ of PBE orbitals and 2-center $J$ \\ 
\begin{tabular}{lll}
0.25 electrons per atom  & & \\ 
\textbf{95.15\%} & \textbf{(1,1,0,0,0,0,0,0)} & \textbf{(0.5,-0.5,0,0,0,0,0,0)} \\ 
4.85\% & (2,0,0,0,0,0,0,0) & (0,0,0,0,0,0,0,0) \\\\ 
0.5 electrons per atom  & & \\ 
\textbf{75.61\%} & \textbf{(1,1,1,1,0,0,0,0)} & \textbf{(0.5,0.5,-0.5,-0.5,0,0,0,0)} \\ 
22.73\% & (2,1,1,0,0,0,0,0) & (0,0.5,-0.5,0,0,0,0,0) \\ 
1.66\% & (2,2,0,0,0,0,0,0) & (0,0,0,0,0,0,0,0) \\\\ 
0.75 electrons per atom  & & \\ 
45.41\% & (2,1,1,1,1,0,0,0) & (0,0.5,0.5,-0.5,-0.5,0,0,0) \\ 
\textbf{39.26\%} & \textbf{(1,1,1,1,1,1,0,0)} & \textbf{(0.5,0.5,0.5,-0.5,-0.5,-0.5,0,0)} \\ 
14.48\% & (2,2,1,1,0,0,0,0) & (0,0,0.5,-0.5,0,0,0,0) \\ 
0.84\% & (2,2,2,0,0,0,0,0) & (0,0,0,0,0,0,0,0) \\ 
0.01\% & (2,1,1,1,1,0,0,0) & (1.0,0.5,-0.5,-0.5,-0.5,0,0,0) \\\\ 
1.0 electrons per atom  & & \\ 
38.87\% & (2,1,1,1,1,1,1,0) & (0,0.5,0.5,0.5,-0.5,-0.5,-0.5,0) \\ 
29.61\% & (2,2,1,1,1,1,0,0) & (0,0,0.5,0.5,-0.5,-0.5,0,0) \\ 
\textbf{19.14\%} & \textbf{(1,1,1,1,1,1,1,1)} & \textbf{(0.5,0.5,0.5,0.5,-0.5,-0.5,-0.5,-0.5)} \\ 
11.04\% & (2,2,2,1,1,0,0,0) & (0,0,0,0.5,-0.5,0,0,0) \\ 
1.34\% & (2,2,2,2,0,0,0,0) & (0,0,0,0,0,0,0,0) \\\\ 
1.25 electrons per atom  & & \\ 
46.17\% & (2,2,2,1,1,1,1,0) & (0,0,0,0.5,0.5,-0.5,-0.5,0) \\ 
\textbf{37.72\%} & \textbf{(2,2,1,1,1,1,1,1)} & \textbf{(0,0,0.5,0.5,0.5,-0.5,-0.5,-0.5)} \\ 
15.14\% & (2,2,2,2,1,1,0,0) & (0,0,0,0,0.5,-0.5,0,0) \\ 
0.98\% & (2,2,2,2,2,0,0,0) & (0,0,0,0,0,0,0,0) \\\\ 
1.5 electrons per atom  & & \\ 
\textbf{68.4\%} & \textbf{(2,2,2,2,1,1,1,1)} & \textbf{(0,0,0,0,0.5,0.5,-0.5,-0.5)} \\ 
28.81\% & (2,2,2,2,2,1,1,0) & (0,0,0,0,0,0.5,-0.5,0) \\ 
2.78\% & (2,2,2,2,2,2,0,0) & (0,0,0,0,0,0,0,0) \\ 
0.01\% & (3,2,2,1,1,1,1,1) & (0.5,0,0,0.5,0.5,-0.5,-0.5,-0.5) \\\\ 
1.75 electrons per atom  & & \\ 
\textbf{91.64\%} & \textbf{(2,2,2,2,2,2,1,1)} & \textbf{(0,0,0,0,0,0,0.5,-0.5)} \\ 
8.34\% & (2,2,2,2,2,2,2,0) & (0,0,0,0,0,0,0,0) \\ 
0.01\% & (3,2,2,2,2,1,1,1) & (0.5,0,0,0,0,0.5,-0.5,-0.5) \\ 
0.01\% & (3,2,2,2,2,1,1,1) & (-0.5,0,0,0,0,0.5,0.5,-0.5) \\\\ 
2.0 electrons per atom  & & \\ 
\textbf{99.6\%} & \textbf{(2,2,2,2,2,2,2,2)} & \textbf{(0,0,0,0,0,0,0,0)} \\ 
0.22\% & (3,2,2,2,2,2,2,1) & (0.5,0,0,0,0,0,0,-0.5) \\ 
0.18\% & (3,2,2,2,2,2,2,1) & (-0.5,0,0,0,0,0,0,0.5) \\\\ 
\end{tabular} 
\label{tableHcubePBEeqj} 
\end{table}

\begin{table}[h!] 
\centering 
H-cube, FNDMC with $\Psi_T=\Psi_S J$ using $\Psi_S$ of PBE orbitals and 2-center $J$ \\ 
\begin{tabular}{lll}
0.25 electrons per atom  & & \\ 
\textbf{100.0\%} & \textbf{(1,1,0,0,0,0,0,0)} & \textbf{(0.5,-0.5,0,0,0,0,0,0)} \\\\ 
0.5 electrons per atom  & & \\ 
\textbf{100.0\%} & \textbf{(1,1,1,1,0,0,0,0)} & \textbf{(0.5,0.5,-0.5,-0.5,0,0,0,0)} \\\\ 
0.75 electrons per atom  & & \\ 
\textbf{100.0\%} & \textbf{(1,1,1,1,1,1,0,0)} & \textbf{(0.5,0.5,0.5,-0.5,-0.5,-0.5,0,0)} \\\\ 
1.0 electrons per atom  & & \\ 
\textbf{100.0\%} & \textbf{(1,1,1,1,1,1,1,1)} & \textbf{(0.5,0.5,0.5,0.5,-0.5,-0.5,-0.5,-0.5)} \\\\ 
1.25 electrons per atom  & & \\ 
\textbf{100.0\%} & \textbf{(2,2,1,1,1,1,1,1)} & \textbf{(0,0,0.5,0.5,0.5,-0.5,-0.5,-0.5)} \\\\ 
1.5 electrons per atom  & & \\ 
\textbf{100.0\%} & \textbf{(2,2,2,2,1,1,1,1)} & \textbf{(0,0,0,0,0.5,0.5,-0.5,-0.5)} \\\\ 
1.75 electrons per atom  & & \\ 
\textbf{100.0\%} & \textbf{(2,2,2,2,2,2,1,1)} & \textbf{(0,0,0,0,0,0,0.5,-0.5)} \\\\ 
2.0 electrons per atom  & & \\ 
\textbf{100.0\%} & \textbf{(2,2,2,2,2,2,2,2)} & \textbf{(0,0,0,0,0,0,0,0)} \\\\ 
\end{tabular} 
\label{tableHcubePBEdmcj} 
\end{table}

\begin{table}[h!] 
\centering 
Cl-cube, VMC with $\Psi_S$ of HF orbitals \\ 
\begin{tabular}{lll}
16.0 electrons per atom  & & \\ 
\textbf{100.0\%} & \textbf{(6,6,6,6,6,6,6,6)} & \textbf{(0,0,0,0,0,0,0,0)} \\\\ 
16.25 electrons per atom  & & \\ 
\textbf{87.92\%} & \textbf{(7,7,6,6,6,6,6,6)} & \textbf{(0.5,-0.5,0,0,0,0,0,0)} \\ 
12.08\% & (8,6,6,6,6,6,6,6) & (0,0,0,0,0,0,0,0) \\\\ 
16.5 electrons per atom  & & \\ 
\textbf{56.93\%} & \textbf{(7,7,7,7,6,6,6,6)} & \textbf{(0.5,0.5,-0.5,-0.5,0,0,0,0)} \\ 
36.99\% & (8,7,7,6,6,6,6,6) & (0,0.5,-0.5,0,0,0,0,0) \\ 
6.08\% & (8,8,6,6,6,6,6,6) & (0,0,0,0,0,0,0,0) \\\\ 
16.75 electrons per atom  & & \\ 
47.27\% & (8,7,7,7,7,6,6,6) & (0,0.5,0.5,-0.5,-0.5,0,0,0) \\ 
27.42\% & (8,8,7,7,6,6,6,6) & (0,0,0.5,-0.5,0,0,0,0) \\ 
\textbf{22.03\%} & \textbf{(7,7,7,7,7,7,6,6)} & \textbf{(0.5,0.5,0.5,-0.5,-0.5,-0.5,0,0)} \\ 
3.28\% & (8,8,8,6,6,6,6,6) & (0,0,0,0,0,0,0,0) \\\\ 
17.0 electrons per atom  & & \\ 
51.11\% & (8,8,7,7,7,7,6,6) & (0,0,0.5,0.5,-0.5,-0.5,0,0) \\ 
23.02\% & (8,7,7,7,7,7,7,6) & (0,0.5,0.5,0.5,-0.5,-0.5,-0.5,0) \\ 
21.77\% & (8,8,8,7,7,6,6,6) & (0,0,0,0.5,-0.5,0,0,0) \\ 
\textbf{2.08\%} & \textbf{(7,7,7,7,7,7,7,7)} & \textbf{(0.5,0.5,0.5,0.5,-0.5,-0.5,-0.5,-0.5)} \\ 
2.03\% & (8,8,8,8,6,6,6,6) & (0,0,0,0,0,0,0,0) \\\\ 
17.25 electrons per atom  & & \\ 
47.56\% & (8,8,8,7,7,7,7,6) & (0,0,0,0.5,0.5,-0.5,-0.5,0) \\ 
27.94\% & (8,8,8,8,7,7,6,6) & (0,0,0,0,0.5,-0.5,0,0) \\ 
\textbf{21.4\%} & \textbf{(8,8,7,7,7,7,7,7)} & \textbf{(0,0,0.5,0.5,0.5,-0.5,-0.5,-0.5)} \\ 
3.1\% & (8,8,8,8,8,6,6,6) & (0,0,0,0,0,0,0,0) \\\\ 
17.5 electrons per atom  & & \\ 
\textbf{56.8\%} & \textbf{(8,8,8,8,7,7,7,7)} & \textbf{(0,0,0,0,0.5,0.5,-0.5,-0.5)} \\ 
36.94\% & (8,8,8,8,8,7,7,6) & (0,0,0,0,0,0.5,-0.5,0) \\ 
6.26\% & (8,8,8,8,8,8,6,6) & (0,0,0,0,0,0,0,0) \\\\ 
17.75 electrons per atom  & & \\ 
\textbf{87.56\%} & \textbf{(8,8,8,8,8,8,7,7)} & \textbf{(0,0,0,0,0,0,0.5,-0.5)} \\ 
12.44\% & (8,8,8,8,8,8,8,6) & (0,0,0,0,0,0,0,0) \\\\ 
18.0 electrons per atom  & & \\ 
\textbf{100.0\%} & \textbf{(8,8,8,8,8,8,8,8)} & \textbf{(0,0,0,0,0,0,0,0)} \\\\ 
\end{tabular} 
\label{tableClcubeRHFeqnoj} 
\end{table}

\begin{table}[h!] 
\centering 
Cl-cube, VMC with $\Psi_T=\Psi_S J$ using $\Psi_S$ of HSE06 orbitals and 2-center $J$ \\ 
\begin{tabular}{lll}
16.0 electrons per atom  & & \\ 
\textbf{100.0\%} & \textbf{(6,6,6,6,6,6,6,6)} & \textbf{(0,0,0,0,0,0,0,0)} \\\\ 
16.25 electrons per atom  & & \\ 
\textbf{87.5\%} & \textbf{(7,7,6,6,6,6,6,6)} & \textbf{(0.5,-0.5,0,0,0,0,0,0)} \\ 
12.5\% & (8,6,6,6,6,6,6,6) & (0,0,0,0,0,0,0,0) \\\\ 
16.5 electrons per atom  & & \\ 
\textbf{55.2\%} & \textbf{(7,7,7,7,6,6,6,6)} & \textbf{(0.5,0.5,-0.5,-0.5,0,0,0,0)} \\ 
38.64\% & (8,7,7,6,6,6,6,6) & (0,0.5,-0.5,0,0,0,0,0) \\ 
6.16\% & (8,8,6,6,6,6,6,6) & (0,0,0,0,0,0,0,0) \\\\ 
16.75 electrons per atom  & & \\ 
46.81\% & (8,7,7,7,7,6,6,6) & (0,0.5,0.5,-0.5,-0.5,0,0,0) \\ 
28.0\% & (8,8,7,7,6,6,6,6) & (0,0,0.5,-0.5,0,0,0,0) \\ 
\textbf{21.86\%} & \textbf{(7,7,7,7,7,7,6,6)} & \textbf{(0.5,0.5,0.5,-0.5,-0.5,-0.5,0,0)} \\ 
3.32\% & (8,8,8,6,6,6,6,6) & (0,0,0,0,0,0,0,0) \\\\ 
17.0 electrons per atom  & & \\ 
50.22\% & (8,8,7,7,7,7,6,6) & (0,0,0.5,0.5,-0.5,-0.5,0,0) \\ 
24.24\% & (8,8,8,7,7,6,6,6) & (0,0,0,0.5,-0.5,0,0,0) \\ 
20.94\% & (8,7,7,7,7,7,7,6) & (0,0.5,0.5,0.5,-0.5,-0.5,-0.5,0) \\ 
2.91\% & (8,8,8,8,6,6,6,6) & (0,0,0,0,0,0,0,0) \\ 
\textbf{1.7\%} & \textbf{(7,7,7,7,7,7,7,7)} & \textbf{(0.5,0.5,0.5,0.5,-0.5,-0.5,-0.5,-0.5)} \\\\ 
17.25 electrons per atom  & & \\ 
45.98\% & (8,8,8,7,7,7,7,6) & (0,0,0,0.5,0.5,-0.5,-0.5,0) \\ 
29.92\% & (8,8,8,8,7,7,6,6) & (0,0,0,0,0.5,-0.5,0,0) \\ 
\textbf{20.69\%} & \textbf{(8,8,7,7,7,7,7,7)} & \textbf{(0,0,0.5,0.5,0.5,-0.5,-0.5,-0.5)} \\ 
3.41\% & (8,8,8,8,8,6,6,6) & (0,0,0,0,0,0,0,0) \\\\ 
17.5 electrons per atom  & & \\ 
\textbf{54.75\%} & \textbf{(8,8,8,8,7,7,7,7)} & \textbf{(0,0,0,0,0.5,0.5,-0.5,-0.5)} \\ 
38.32\% & (8,8,8,8,8,7,7,6) & (0,0,0,0,0,0.5,-0.5,0) \\ 
6.93\% & (8,8,8,8,8,8,6,6) & (0,0,0,0,0,0,0,0) \\\\ 
17.75 electrons per atom  & & \\ 
\textbf{86.5\%} & \textbf{(8,8,8,8,8,8,7,7)} & \textbf{(0,0,0,0,0,0,0.5,-0.5)} \\ 
13.5\% & (8,8,8,8,8,8,8,6) & (0,0,0,0,0,0,0,0) \\\\ 
18.0 electrons per atom  & & \\ 
\textbf{100.0\%} & \textbf{(8,8,8,8,8,8,8,8)} & \textbf{(0,0,0,0,0,0,0,0)} \\\\ 
\end{tabular} 
\label{tableClcubeHSEH1PBEeqj} 
\end{table}

\begin{table}[h!] 
\centering 
Cl-cube, VMC with $\Psi_T=\Psi_S J$ using $\Psi_S$ of HF orbitals and 2-center $J$ \\ 
\begin{tabular}{lll}
16.0 electrons per atom  & & \\ 
\textbf{100.0\%} & \textbf{(6,6,6,6,6,6,6,6)} & \textbf{(0,0,0,0,0,0,0,0)} \\\\ 
16.25 electrons per atom  & & \\ 
\textbf{87.98\%} & \textbf{(7,7,6,6,6,6,6,6)} & \textbf{(0.5,-0.5,0,0,0,0,0,0)} \\ 
12.02\% & (8,6,6,6,6,6,6,6) & (0,0,0,0,0,0,0,0) \\\\ 
16.5 electrons per atom  & & \\ 
\textbf{54.83\%} & \textbf{(7,7,7,7,6,6,6,6)} & \textbf{(0.5,0.5,-0.5,-0.5,0,0,0,0)} \\ 
39.07\% & (8,7,7,6,6,6,6,6) & (0,0.5,-0.5,0,0,0,0,0) \\ 
6.09\% & (8,8,6,6,6,6,6,6) & (0,0,0,0,0,0,0,0) \\\\ 
16.75 electrons per atom  & & \\ 
45.98\% & (8,7,7,7,7,6,6,6) & (0,0.5,0.5,-0.5,-0.5,0,0,0) \\ 
30.3\% & (8,8,7,7,6,6,6,6) & (0,0,0.5,-0.5,0,0,0,0) \\ 
\textbf{19.91\%} & \textbf{(7,7,7,7,7,7,6,6)} & \textbf{(0.5,0.5,0.5,-0.5,-0.5,-0.5,0,0)} \\ 
3.81\% & (8,8,8,6,6,6,6,6) & (0,0,0,0,0,0,0,0) \\\\ 
17.0 electrons per atom  & & \\ 
50.23\% & (8,8,7,7,7,7,6,6) & (0,0,0.5,0.5,-0.5,-0.5,0,0) \\ 
22.97\% & (8,8,8,7,7,6,6,6) & (0,0,0,0.5,-0.5,0,0,0) \\ 
22.49\% & (8,7,7,7,7,7,7,6) & (0,0.5,0.5,0.5,-0.5,-0.5,-0.5,0) \\ 
\textbf{2.26\%} & \textbf{(7,7,7,7,7,7,7,7)} & \textbf{(0.5,0.5,0.5,0.5,-0.5,-0.5,-0.5,-0.5)} \\ 
2.05\% & (8,8,8,8,6,6,6,6) & (0,0,0,0,0,0,0,0) \\\\ 
17.25 electrons per atom  & & \\ 
46.13\% & (8,8,8,7,7,7,7,6) & (0,0,0,0.5,0.5,-0.5,-0.5,0) \\ 
30.42\% & (8,8,8,8,7,7,6,6) & (0,0,0,0,0.5,-0.5,0,0) \\ 
\textbf{20.21\%} & \textbf{(8,8,7,7,7,7,7,7)} & \textbf{(0,0,0.5,0.5,0.5,-0.5,-0.5,-0.5)} \\ 
3.23\% & (8,8,8,8,8,6,6,6) & (0,0,0,0,0,0,0,0) \\\\ 
17.5 electrons per atom  & & \\ 
\textbf{54.27\%} & \textbf{(8,8,8,8,7,7,7,7)} & \textbf{(0,0,0,0,0.5,0.5,-0.5,-0.5)} \\ 
38.65\% & (8,8,8,8,8,7,7,6) & (0,0,0,0,0,0.5,-0.5,0) \\ 
7.08\% & (8,8,8,8,8,8,6,6) & (0,0,0,0,0,0,0,0) \\\\ 
17.75 electrons per atom  & & \\ 
\textbf{86.9\%} & \textbf{(8,8,8,8,8,8,7,7)} & \textbf{(0,0,0,0,0,0,0.5,-0.5)} \\ 
13.1\% & (8,8,8,8,8,8,8,6) & (0,0,0,0,0,0,0,0) \\\\ 
18.0 electrons per atom  & & \\ 
\textbf{100.0\%} & \textbf{(8,8,8,8,8,8,8,8)} & \textbf{(0,0,0,0,0,0,0,0)} \\\\ 
\end{tabular} 
\label{tableClcubeRHFeqj} 
\end{table}

\begin{table}[h!] 
\centering 
Cl-cube, FNDMC with $\Psi_T=\Psi_S J$ using $\Psi_S$ of HSE06 orbitals and 2-center $J$ \\ 
\begin{tabular}{lll}
16.0 electrons per atom  & & \\ 
\textbf{100.0\%} & \textbf{(6,6,6,6,6,6,6,6)} & \textbf{(0,0,0,0,0,0,0,0)} \\\\ 
16.25 electrons per atom  & & \\ 
\textbf{100.0\%} & \textbf{(7,7,6,6,6,6,6,6)} & \textbf{(0.5,-0.5,0,0,0,0,0,0)} \\\\ 
16.5 electrons per atom  & & \\ 
\textbf{100.0\%} & \textbf{(7,7,7,7,6,6,6,6)} & \textbf{(0.5,0.5,-0.5,-0.5,0,0,0,0)} \\\\ 
16.75 electrons per atom  & & \\ 
\textbf{100.0\%} & \textbf{(7,7,7,7,7,7,6,6)} & \textbf{(0.5,0.5,0.5,-0.5,-0.5,-0.5,0,0)} \\\\ 
17.0 electrons per atom  & & \\ 
\textbf{100.0\%} & \textbf{(7,7,7,7,7,7,7,7)} & \textbf{(0.5,0.5,0.5,0.5,-0.5,-0.5,-0.5,-0.5)} \\\\ 
17.25 electrons per atom  & & \\ 
\textbf{100.0\%} & \textbf{(8,8,7,7,7,7,7,7)} & \textbf{(0,0,0.5,0.5,0.5,-0.5,-0.5,-0.5)} \\\\ 
17.5 electrons per atom  & & \\ 
\textbf{100.0\%} & \textbf{(8,8,8,8,7,7,7,7)} & \textbf{(0,0,0,0,0.5,0.5,-0.5,-0.5)} \\\\ 
17.75 electrons per atom  & & \\ 
\textbf{100.0\%} & \textbf{(8,8,8,8,8,8,7,7)} & \textbf{(0,0,0,0,0,0,0.5,-0.5)} \\\\ 
18.0 electrons per atom  & & \\ 
\textbf{100.0\%} & \textbf{(8,8,8,8,8,8,8,8)} & \textbf{(0,0,0,0,0,0,0,0)} \\\\ 
\end{tabular} 
\label{tableClcubeHSEH1PBEdmcj} 
\end{table}

\begin{table}[h!] 
\centering 
Cl-cube, FNDMC with $\Psi_T=\Psi_S J$ using $\Psi_S$ of HF orbitals and 2-center $J$ \\ 
\begin{tabular}{lll}
16.0 electrons per atom  & & \\ 
\textbf{100.0\%} & \textbf{(6,6,6,6,6,6,6,6)} & \textbf{(0,0,0,0,0,0,0,0)} \\\\ 
16.25 electrons per atom  & & \\ 
\textbf{100.0\%} & \textbf{(7,7,6,6,6,6,6,6)} & \textbf{(0.5,-0.5,0,0,0,0,0,0)} \\\\ 
16.5 electrons per atom  & & \\ 
\textbf{100.0\%} & \textbf{(7,7,7,7,6,6,6,6)} & \textbf{(0.5,0.5,-0.5,-0.5,0,0,0,0)} \\\\ 
16.75 electrons per atom  & & \\ 
\textbf{100.0\%} & \textbf{(7,7,7,7,7,7,6,6)} & \textbf{(0.5,0.5,0.5,-0.5,-0.5,-0.5,0,0)} \\\\ 
17.0 electrons per atom  & & \\ 
\textbf{100.0\%} & \textbf{(7,7,7,7,7,7,7,7)} & \textbf{(0.5,0.5,0.5,0.5,-0.5,-0.5,-0.5,-0.5)} \\\\ 
17.25 electrons per atom  & & \\ 
\textbf{100.0\%} & \textbf{(8,8,7,7,7,7,7,7)} & \textbf{(0,0,0.5,0.5,0.5,-0.5,-0.5,-0.5)} \\\\ 
17.5 electrons per atom  & & \\ 
\textbf{100.0\%} & \textbf{(8,8,8,8,7,7,7,7)} & \textbf{(0,0,0,0,0.5,0.5,-0.5,-0.5)} \\\\ 
17.75 electrons per atom  & & \\ 
\textbf{100.0\%} & \textbf{(8,8,8,8,8,8,7,7)} & \textbf{(0,0,0,0,0,0,0.5,-0.5)} \\\\ 
18.0 electrons per atom  & & \\ 
\textbf{100.0\%} & \textbf{(8,8,8,8,8,8,8,8)} & \textbf{(0,0,0,0,0,0,0,0)} \\\\ 
\end{tabular} 
\label{tableClcubeRHFdmcj} 
\end{table}

}

\end{document}